\documentclass[final,authoryear,5p]{elsarticle}
\pdfoutput=1
\newcommand{\apj}{ApJ}
\newcommand{\apjl}{ApJ Lett.}
\newcommand{\apjs}{ApJS}
\newcommand{\aj}{AJ}
\newcommand{\aap}{A\&A}
\newcommand{\mnras}{MNRAS}
\newcommand{\nat}{Nature}
\newcommand{\sun}{\odot}
\newcommand{\earth}{\oplus}
\newcommand{\url}{}
\newcommand{\software}{\emph{Software}:~}
\newcommand{\acknowledgments}{\vspace{0.1truein}}

 \usepackage{epsfig}

\usepackage{amssymb}

\journal{Advances in Space Research}

\begin{document}

%%%%%%%%%%%%%%%%%%%%%%%%%%%%%%%%%%%%%%%%%%%%%%%%%%%%%%%%%%%%%%%%%%%%%%%%%%%%%
%% Frontmatter
\begin{frontmatter}


%\title{Imaging Black Holes and Jets with Space-Based Telescopes}
\title{Imaging Black Holes and Jets with a VLBI Array Including Multiple Space-Based Telescopes}

\author[label1]{Vincent L.\ Fish\corref{cor}}
\cortext[cor]{Corresponding author}
\ead{vfish@haystack.mit.edu}

\author[label1,label2]{Maura Shea}

\author[label1,label3]{Kazunori Akiyama}

\address[label1]{Massachusetts Institute of Technology, Haystack
  Observatory, 99 Millstone Hill Road, Westford, MA 01886, USA}
\address[label2]{Wellesley College, Whitin Observatory, Wellesley, MA
  02482, USA}
\address[label3]{National Radio Astronomy Observatory Jansky Fellow}

\begin{abstract}
  Very long baseline interferometry (VLBI) from the ground at
  millimeter wavelengths can resolve the black hole shadow around two
  supermassive black holes, Sagittarius~A* and M87.  The addition of
  modest telescopes in space would allow the combined array to produce
  higher-resolution, higher-fidelity images of these and other
  sources.  This paper explores the potential benefits of adding
  orbital elements to the Event Horizon Telescope.  We reconstruct
  model images using simulated data from arrays including telescopes
  in different orbits.  We find that an array including one telescope
  near geostationary orbit and one in a high-inclination medium Earth
  or geosynchronous orbit can succesfully produce high-fidelity images
  capable of resolving shadows as small as $3~\mu$as in diameter.  One
  such key source, the Sombrero Galaxy, may be important to address
  questions regarding why some black holes launch powerful jets while
  others do not.  Meanwhile, higher-resolution imaging of the
  substructure of M87 may clarify how jets are launched in the first
  place.  The extra resolution provided by space VLBI will also
  improve studies of the collimation of jets from active galactic
  nuclei.
\end{abstract}


\end{frontmatter}

\parindent=0.5 cm

%%%%%%%%%%%%%%%%%%%%%%%%%%%%%%%%%%%%%%%%%%%%%%%%%%%%%%%%%%%%%%%%%%%%%%%%%%%%%
%% Main text

\copyright\ 2019.  This manuscript version is made available under the CC-BY-NC-ND 4.0 license\footnote{\url{http://creativecommons.org/licenses/by-nc-nd/4.0/}}

\section{Introduction}

The angular resolution of an interferometric baseline is approximately
the observing wavelength divided by the baseline length $\lambda/B$.
The choice of observing wavelength is often fixed by source
properties, and in any case atmospheric absorption imposes
site-dependent limits on what is possible.  Many terrestrial arrays
using very long baseline interferometry (VLBI), such as the Very Long
Baseline Array (VLBA) or the Event Horizon Telescope (EHT), achieve
high angular resolution by having telescopes that are thousands of
kilometers apart.  Here too, the size of the Earth imposes fundamental
limits on the longest achievable baseline from the ground.

Longer baselines are achievable with space-borne elements.  Space VLBI
has successfully been accomplished at centimeter wavelengths with the
Tracking and Data Relay Satellite System \citep[TDRSS;][]{levy1986},
Highly Advance Laboratory for Communications and Astronomy (HALCA)
VLBI Space Observatory Programme \citep[VSOP;][]{hirabayashi1998}, and
RadioAstron \citep{kardashev2013}.  Arrays with two space-borne
elements have been proposed before at frequencies up to 43~GHz and
86~GHz \citep{hong2004,murphy2005}.

The EHT is a millimeter-wavelength VLBI array whose primary goal is to
image nearby supermassive black holes and jets of active galactic
nuclei (AGNs).  Currently observing at $\lambda = 1.3$~mm (230~GHz),
the fringe spacing ($\lambda/B$, where $B$ is the projected baseline
length) of the longest baselines correspond to an angular resolution
$\lesssim 25~\mu$as, which is sufficient to resolve the shadow of the
supermassive black holes of Sagittarius~A* and M87.  The EHT is
currently upgrading telescopes to perform VLBI at $\lambda = 0.87$~mm
(345~GHz), which will further improve angular resolution by a factor
of 1.5.

The EHT has developed imaging algorithms to produce superresolved
images to make the most out of its data
\citep[e.g.,][]{bouman2016,chael2016,akiyama2017a,akiyama2017b,kuramochi2018}.  Nevertheless, the
EHT is up against some hard limits.  The Earth's atmosphere quickly
becomes unsuitable for ground-based VLBI at higher frequencies, and
existing baselines already approach an Earth diameter.  In order to
achieve higher angular resolution, it will be necessary to incorporate
space-borne elements into VLBI arrays of the future.

Adding space-borne antennas to the EHT opens up new science that is
very difficult or impossible from the ground.  Reconstructing reliable
movies of Sgr~A* is likely to require fast $(u,v)$ coverage not
currently obtainable from Earth-rotation aperture
synthesis\footnote{It is possible that a large number of additional
  ground stations could enable snapshot imaging of Sgr~A*, although
  the geographical distribution of sites suitable for 230~GHz
  observing imposes fundamental limits on a ground-only approach.}
\citep{palumbo2018}.  The ability to resolve and image fine-scale
structures (smaller than the gravitational radius, $r_g = GM/c^2$) in
the flow around M87 will help enormously in understanding the details
of how jets are launched.  Adding to the EHT one or more telescopes
near geosynchronous orbit would enable detailed study of other black
hole sources such as the Sombrero Galaxy, an M87 analogue with a much
weaker jet.  Space VLBI will also help answer the question of whether
jet collimation is universal across a wide range of jet power.  And,
as is clear from the diversity of presentations at the workshop
``The Future of High-Resolution Radio Interferometry in Space'' held
in Noordwijk, The Netherlands, in
2018\footnote{\url{https://www.ru.nl/astrophysics/news-agenda/future-high-resolution-radio-interferometry-space/submitted-abstracts/} (accessed 2019 March 13)},
AGN jet science is just one of the many areas where space VLBI could
have a large impact.

In this work, we focus on the applicability of a 230~GHz space-VLBI
array to AGN jet studies.  We motivate an architecture and possible
space-VLBI array from practical considerations
(Section~\ref{considerations}), generate synthetic data
(Section~\ref{methods}), and examine the imaging power of such an
array (Section~\ref{results}).  We find that an array consisting of a
half dozen or so satellites of modest aperture can provide the fast
microarcsecond-scale angular resolution needed to make next-generation
breakthroughs in AGN jet science.

\section{Considerations for Designing a Space VLBI Array} \label{considerations}

\subsection{Technical Assumptions}\label{assumptions}

This work focuses on what could be achieved scientifically by
launching highly capable satellites into several classes of orbits
around the Earth.  Issues regarding technology and engineering are
necessarily beyond the scope of this work, and in any case it is
impossible to predict with full accuracy what the landscape will look
like in the future.  Nevertheless, it is worth specifying a few key
technical assumptions to assess whether the concept described in this
paper might be feasible within the next 5--10 years.

The cost of access to space is decreasing.  Rideshare opportunities
are becoming plentiful, with reduced or even zero marginal launch
costs for secondary payloads.  Taking full advantage of these
opportunities will likely require that space-VLBI payloads fit within
the size and weight limits of a small satellite that could be launched
from an EELV (Evolved Expendable Launch Vehicles) Secondary Payload
Adapter (ESPA) Grande ring, for instance.  As an added benefit,
keeping the payload size small may reduce the cost per satellite,
potentially allowing for more satellites to be built and launched.  A
rideshare strategy argues for choosing classes of orbits with frequent
launches (e.g., LEO and GEO) rather than requiring bespoke orbital
parameters.

Our vision assumes improvements in some space-borne telescope
subsystems that we believe to be tractable within less than a decade.
We assume that an aperture of a few meters in diameter with
appropriate surface accuracy for observations at 1.3~mm can be
deployed cheaply.  One promising approach might be to stow the surface
within a standard secondary payload volume and unfurl it in space.
Other approaches may become financially viable as the increasing
number of commercial launch opporunities drive costs down.  We assume
that stable receivers and signal chains with bandwidths of many GHz
can provide adequate sensitivity at a cost that is not prohibitive.
We assume that a very accurate frequency standard can be provided at
reasonable cost.  Onboard atomic clocks on each element may be the
easiest solution, although distributed signals with a round-trip loop
may also be feasible.  We assume that laser communications links will
be able to support data rates of many gigabits per second.
Transferring large amounts of data to the ground may require
substantial onboard data storage and a geosynchronous satellite acting
as a relay.

We have been purposefully vague in the preceding paragraph.  Other
scientific, industrial, and military applications are already driving
some of the assumed advances.  Given adequate time and funding,
focused efforts could enable major progress in the remaining areas.
All of the required pieces of technology for a space-VLBI 1.3~mm array
exist, even if cost or performance issues may preclude advancing a
complete mission today.  Furthermore, as will become clearer in
Section~\ref{sensitivity}, the achievable performance of a space-VLBI
array depends on multiple parameters (aperture size, aperture
efficiency, bandwidth, and system temperature) of both the space-based
and ground-based elements.  Improvements in one of these parameters
are interchangeable with improvements in another.

\subsection{Timescales}\label{timescales}

In order to produce static images or movies of a source with varying
structure, it is desirable to sample the $(u,v)$ plane before the
source structure changes appreciaby.  A useful characteristic
timescale for a black hole source is $t_g = GM/c^3$, the
light-crossing time of the gravitational radius.  For Sgr A*, whose
mass is approximately $4.3 \times 10^6~M_\sun$, this time is about
20~s.  Structures in an accretion flow may be bigger than $r_g$, and
material in the accretion flow moves at subluminal velocities, so a
source may effectively be static over a timescale of a few $t_g$.
Indeed, Sgr~A* is seen to vary across the electromagnetic spectrum on
timescales of minutes \citep[e.g.,][]{marrone2008}.

Other supermassive black holes of interest are much more massive and
therefore vary on longer timescales.  For instance, $t_g \approx 9$~hr
for M87, assuming a mass of $6.6 \times 10^9~M_\sun$
\citep{gebhardt2011}.  For these sources, visibilities obtained over the
course of a day are sampling a nearly static source.

For more distant sources, structural changes are only relevant if they
are on a sufficiently large angular scale to be detected.  AGN jets
often exhibit superluminal apparent motion, but these sources are
farther away, resulting in a small apparent angular motion on the sky.
For instance, features in the jet of 3C~279 are seen to move at many
times the speed of light \citep[][and many
  others]{whitney1971,kellermann1974,cohen1977}, corresponding to a
motion of $1~\mu$as in a few days.  Such structural changes are
evident in early EHT data, which detected small changes in the closure
phase even on a triangle of stations with a longest baseline of
$3$--$4$~G$\lambda$ \citep{lu2013}.  The fractional changes in the
data on longer (e.g., space--ground) baselines would, of course, be
substantially larger.

It is possible for small portions of a source to vary in brightness on
even faster timescales.  In this case, it may be possible to mitigate
the loss in image fidelity by treating the constant and variable
components separately when calibrating the data, as is sometimes done
for connected-element interferometry of Sgr~A*
\citep[e.g.,][]{marrone2007,marrone2008}.

Thus, with the exception of Sgr A*, nearly all supermassive black hole
targets of a millimeter-wavelength space-VLBI array can be considered
to be static on the timescale of a day.  The orbits of the satellites
in a millimeter-wavelength space-VLBI array should therefore be
designed to swing through $(u,v)$ space on timescales of approximately
one day or less.  This argues for the highest element of an
Earth-centered space-VLBI array to be near geosynchronous orbit.

\subsection{Baseline Coverage}

All things being equal, a VLBI array with smaller coverage holes in
the $(u,v)$ plane will produce images with higher fidelity, since
there are fewer missing data points for image reconstruction
algorithms to have to try to fill in.  The finer angular resolution
provided by very long baselines is desirable, but it can be difficult
or impossible to reconstruct images from long-baseline data without
data on short and intermediate baselines to fill in the gaps.  This
argues against placing a single satellite in a very high orbit.

The short baselines will be especially important in all but the most
compact sources.  A baseline of $1~D_\earth$ is approximately
$10~$G$\lambda$ at $\lambda = 1.3$~mm, with a fringe spacing of
$\lambda/D \approx 20~\mu$as.  Almost all AGN jet sources have
structure on scales larger than this, sometimes into the hundreds or
thousands of microarcseconds in extent.  Short ($< 1~D_\earth$)
baselines to reconstruct the large-scale emission will be necessary to
make use of longer baselines for the increased angular resolution.
This argues for the inclusion of either ground--ground VLBI baselines
or enough elements in low Earth orbit (LEO) to fill in the center of
the $(u,v)$ plane.

Sky images are inherently two-dimensional.  Previous space-VLBI
efforts have focused on placing a single element into orbit
\citep{levy1986,hirabayashi1998,kardashev2013}.  RadioAstron provides
an instructive case: its images often suffer from having poor
resolution orthogonal to the direction of its orbit
\citep[e.g.,][]{gomez2016}.  An imaging array should therefore have at
least two elements in high, approximately orthogonal orbits.

\subsection{Classes of Orbits}\label{orbittypes}

Low Earth orbits range from a few hundred to 2000~km above the ground.
Since the mean radius of the Earth is approximately 6370~km,
satellites in LEO do not significantly extend angular resolution
beyond what is available from a ground-based array alone.  However,
LEO orbits provide fast baseline coverage.  Satellites in LEO circle
the Earth in approximately 90 to 120 minutes.  LEO--ground and
LEO--LEO baselines can quickly fill in the $(u,v)$ plane out to $\sim
12$~G$\lambda$ at 1.3~mm.  Even for ``snapshot'' observations of a few
minutes, during which ground--ground baselines are effectively
stationary in the $(u,v)$ plane, LEO--ground baselines sweep out
substantial arcs.  Satellites in LEO may therefore be essential for
dynamic imaging of Sgr~A* \citep{palumbo2018}.  Satellites in LEO are
also helpful for fast imaging of M87, though they are not as critical
as for Sgr~A*, due to the longer timescale of variability in M87
(Sec.~\ref{timescales}).

With orbits at $\sim 6.6~R_\earth$, satellites in geosynchronous Earth
orbit (GEO) provide significantly more angular resolution than LEO.
More than one satellite near GEO or Medium Earth orbit (MEO) may be
required in order to provide approximately equal angular resolution at
a wide range of position angles on the sky.  A GEO satellite could
serve double duty as the communications link to other satellites in
the array.  Satellites in LEO are only visible from a ground station
for a small fraction of their orbit, yet for most of their orbit they
have a direct line of sight to a geosynchronous satellite, which in
turn is always visible from the ground.

Higher orbits provide even greater angular resolution at the expense
of slower $(u,v)$ coverage.  Very high orbits (e.g., translunar or
Sun--Earth L2) may provide insufficient $(u,v)$ coverage to produce a
high-fidelity image within the timescale of variation of many sources,
with the additional problem that the lack of intermediate-length
baselines would make it very difficult to connect data from the remote
space-borne element to the ground array (or satellites near the
Earth).  Therefore, for the remainder of this work we consider space
arrays consisting only of telescopes in GEO or lower: GEO or high MEO
satellites for angular resolution, LEO satellites for fast baseline
coverage (for snapshot imaging of Sgr A*) or dense sampling of
baselines within about an Earth diameter (for higher image fidelity of
other sources), and ground-based telescopes for sensitivity.

\subsection{Aperture and Sensitivity}\label{sensitivity}

It can be expensive to build and launch large apertures.  Rocket
payload fairings will impose a maximum upper size to potential
satellites.  While we cannot predict at this time what this size will
be many years in the future, it seems prudent to try to limit the
aperture size to a few meters in diameter.

The system equivalent flux density (SEFD) of a telescope is related to
the geometric area ($A$), aperture efficiency ($\eta_A$), and the
system temperature ($T_\mathrm{sys}$) by
\begin{eqnarray}
  \mathrm{SEFD} &=& \frac{2kT_\mathrm{sys}}{\eta_{A}A} \nonumber \\
  &\approx&
42000~\mathrm{Jy} \left(\frac{D}{3~\mathrm{m}}\right)^{-2}
\left(\frac{\eta_A}{0.7}\right)^{-1}
\left(\frac{T_\mathrm{sys}}{75~\mathrm{K}}\right),
\end{eqnarray}
with lower values corresponding to more sensitive telescopes.  As a
point of reference, the system temperature of the Atacama Large
Millimeter/submillimeter Array (ALMA) is approximately 75~K in the
lower part of the 230~GHz band in very dry conditions
\citep{warmels2018}.  While all current ground-based observatories in
the EHT are capable of producing a lower SEFD at high elevation in
good weather, some EHT scans in 2013 with the Submillimeter Telescope
(SMT) in Arizona or a single Combined Array for Research in
Millimeter-wave Astronomy (CARMA) dish in California had SEFDs near or
above 42000~Jy on Sgr~A* due to a combination of mediocre weather and
low source elevation \citep{johnson2015}.  A satellite of modest
aperture can achieve the sensitivity of a significantly larger
aperture on the ground.

The sensitivity of a baseline of two telescopes is given by
\begin{equation}
  \sigma = \frac{1}{\eta_Q}
  \sqrt{\frac{\mathrm{SEFD}_1 \mathrm{SEFD}_2}{2 \Delta\nu \tau}},
\end{equation}
where $\sigma$ is the noise, $\eta_Q$ is the sampling quantization
loss factor ($\sim 0.88$ for 2-bit sampling), $\Delta\nu$ is the bandwidth,
and $\tau$ is the integration time.  Thus, all else being equal, the
rms noise on a space--ground baseline scales as
\begin{equation}\label{tradeoff}
\sigma \propto \left(A_\mathrm{space} A_\mathrm{ground} \Delta\nu
\tau\right)^{-1/2},
\end{equation}
where $A$ is the area of each aperture.  This immediately suggests
that a cost-effective strategy to achieve high sensitivity is to build
modest space apertures that leverage the very sensitive apertures on
the ground and to observe with a high bandwidth.  If the partner
station is phased ALMA with an SEFD of 70~Jy \citep{app2013} and an
observing bandwidth of 8~GHz, $\sigma \approx 2$~mJy in one minute to
a 42000~Jy space telescope.  Even wider bandwidths are possible from
ground-based telescopes, as demonstrated by the upgraded Submillimeter
Array (SMA) system \citep{primiani2016}.

These bandwidths are significantly larger than used by previous space
VLBI efforts.  VSOP and RadioAstron both typically observe(d) two
16~MHz channels for a total data rate of 128--144 Mb\,s$^{-1}$,
transmitted back to the ground at radio frequencies
\citep{hirabayashi1998,kardashev2013}.  Both of these missions were
designed in an era when this was a typical data rate for terrestrial
VLBI.

There are many sensitive ground stations, including phased ALMA, the
Large Millimeter Telescope (LMT) Alfonso Serrano in Mexico, the phased
SMA in Hawai`i, and the phased Northern Extended Millimeter Array
(NOEMA) in France.  Any one of these would be likely be sufficient to
obtain rapid detections to a space aperture.  So long as fringes can
be found to all telescopes from a sensitive ground station, data on
all baselines can be fringe fitted and used, even if the instantaneous
signal-to-noise ratio (SNR) is small in any given interval.

\section{Methods} \label{methods}

\subsection{General Approach}

Current studies have created arrays with space-based telescopes that
are designed for one particular purpose.  For instance,
\citet{palumbo2018talk} explored an array consisting of four
telescopes in LEO, optimized for dynamical imaging.
\citet{roelofs2018} explored a minimal configuration of two MEO
satellites in a configuration such that the orbits slowly evolve over
the course of six months to image Sgr~A*.  The uniting theme in these
studies is that they have focused on a single scientific case and
identified a single type of array to address that specific case.  In
this paper, we focus on an array configuration that can flexibly
address multiple science cases.

Building off of the concept of \citet{palumbo2018talk}, we start with
a LEO array consisting of up to four telescopes, which provides
excellent $(u,v)$ coverage within $\sim 10$~G$\lambda$ at 230~GHz.  We
add a geosynchronous satellite to significantly increase the
resolution in the east-west direction.  This single satellite can also
provide north-south coverage for high-declination sources.  To add
north-south coverage for the rest of the sky, we add a satellite in an
inclined, eccentric MEO orbit.  Given these constraints, we have
chosen satellites from among existing unclassified NORAD two-line
elements (TLEs) available from CelesTrak\footnote{satellite numbers
  07276, 19822, 25635, 27854, 29107, and 43132 from
  \url{https://www.celestrak.com/NORAD/elements}} in order to simulate
data from satellites in real orbits.  The selection of these
satellites is quasi-arbitrary, and our results do not appear to be
particularly dependent upon the specific satellites chosen, which is
encouraging for a rideshare concept.  A more detailed trade study of
orbital parameters would, of course, be necessary before a specific
mission is advanced.

\subsection{Imaging}

Representative model images were selected, and then simulated data
were produced using the \texttt{ehtim (eht-imaging)} library.  The
ground array was assumed to include the observatories that currently
observe as part of the EHT: ALMA, the Atacama Pathfinder Experiment
(APEX), the Greenland Telescope (GLT), the James Clark Maxwell
Telescope (JCMT), the LMT, the IRAM 30-meter telescope on Pico Veleta,
the South Pole Telescope (SPT), the SMA, and the SMT.  Other ground
observatories, NOEMA and the ALMA prototype antenna at Kitt Peak, are
currently being upgraded to join the EHT and were therefore included
as well.

To reduce the number of data points to be imaged (thereby speeding
convergence), we simulate data with $\tau = 360$~s and $B = 4$~GHz
over a 24-hour period.  From equation~(\ref{tradeoff}), this is
equivalent to $\tau = 90$~s with two orthogonal polarizations, each of
$B = 8$~GHz.  Large integration times limit the field of view that can
be imaged, although data can be segmented to much shorter time
intervals after fringes are found.  Data with SNR less than 3 were
flagged.

\begin{figure}
  \resizebox{\hsize}{!}{\includegraphics{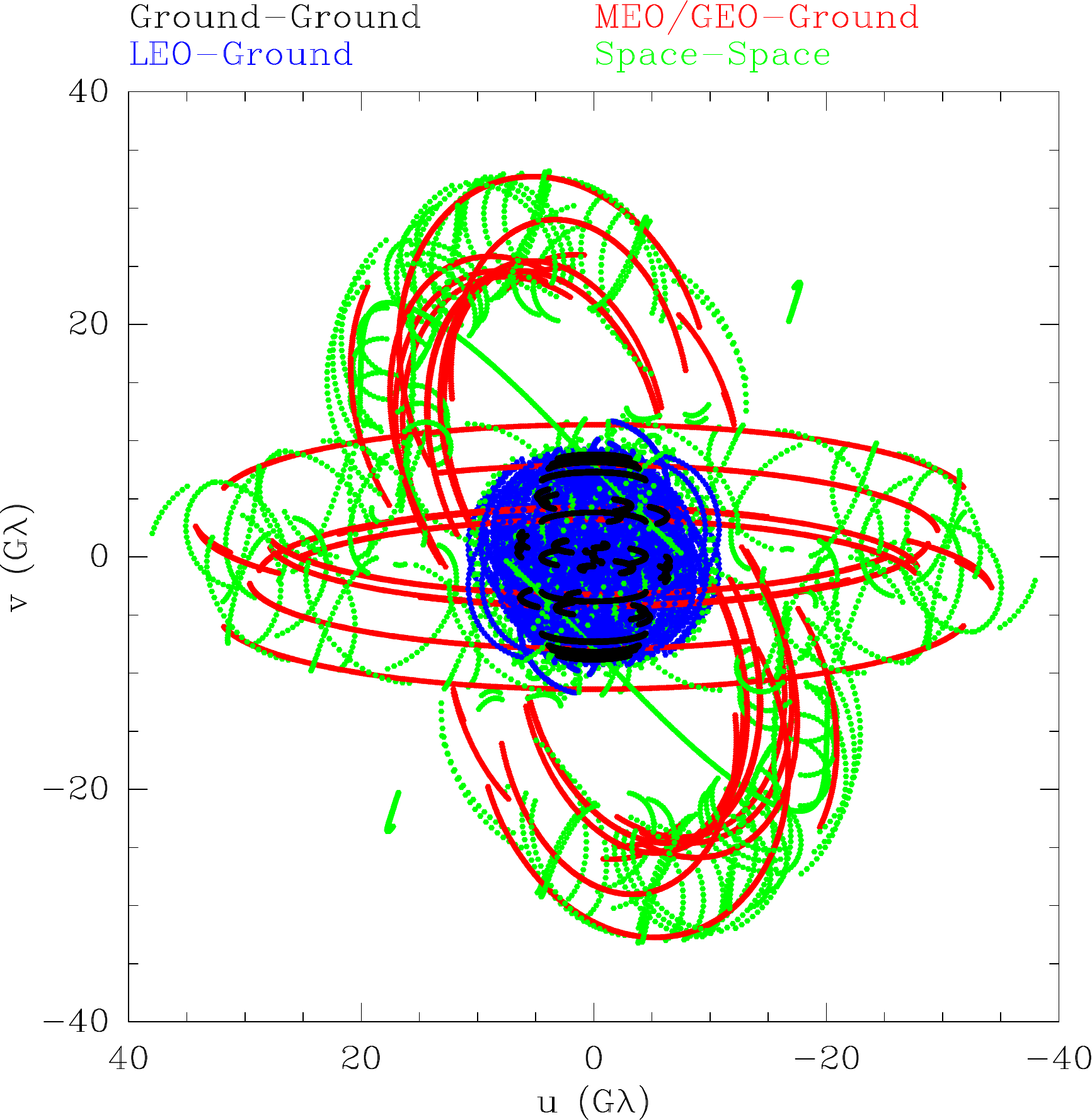}}
  \caption{Representative $(u,v)$ coverage obtainable in 24~hr for a
    source at the declination of M104.  Individual points are spaced
    90~s apart.  The addition of telescopes in MEO/GEO significantly
    expands the maximum baseline length compared to LEO
    alone.}\label{uv-figure}
\end{figure}

Figure~\ref{uv-figure} illustrates a representative $(u,v)$ coverage
that the space array might obtain.  The baselines involving at least
one ground station (red, black, and blue points) have enough
sensitivity to be able to detect sources with as little as a few mJy
of correlated flux density.  For weak sources, the space--space
baselines (green points) may fall below the SNR cutoff for useful
data.  Regardless, the space--space baselines add little $(u,v)$
coverage that cannot be obtained from space--ground baselines
alone\footnote{A possibly significant exception to this is the
  baseline between the two MEO/GEO satellites, which samples a
  different area of $(u,v)$ space.  This baseline vector changes very
  slowly, and it may be possible to integrate for $\tau \gg $ several
  minutes to obtain robust detections.}.

Datasets to image consisted of visibility amplitudes and closure
phases.  For ground-based stations, absolute visibility phases are
difficult to estimate due to very rapid variations in tropospheric
delays.  Space--space baseline phases will be uncontaminated by
these atmospheric contributions, and it is possible that visibility
phases will be usable on these baselines directly, along with hybrid
mapping techniques to estimate visibility phases on other baselines.
Nevertheless, for simplicity our reconstructions use closure phases as
the only phase information included.

Images were then reconstructed from the simulated data using the
Sparse Modeling Imaging Library for Interferometry (SMILI).  In
addition to incorporating a sparsity ($\ell_1$-norm minimizing)
regularizer, SMILI includes total variation (TV) and total square
variation (TSV) regularizers for smoothness
\citep{akiyama2017b,kuramochi2018}.  In our simulations, we use both
the sparsity and TSV regularizers.  Optimal values of the
hyperparameters are determined using a cross-validation approach on a
ground-truth data set.  Images were reconstructed from three arrays: a
ground-only array (i.e., without any elements in space), a ground+LEO
array, and a full array consisting of the ground+LEO array plus one
satellite each in equatorial GEO and high-inclination MEO.

\section{Results} \label{results}

To demonstrate the power of a full space-VLBI array, we examine three
scientific use cases relevant to supermassive black holes and AGN
jets.  Can a full space array resolve the black hole shadow in sources
other than Sgr~A* and M87, and, if so, what is the limit?  Can such an
array resolve details of the jet launch zone of M87?  And can it bring
out the fine details necessary to help understand the collimation and
propagation of AGN jets?

\subsection{Resolving Shadows Around Other Black Holes}

The two prime targets of the EHT, Sgr~A* and M87, are the only known
black hole sources for which terrestrial VLBI at 1.3~mm can resolve
the shadow around the black hole.  The increased resolution of space
VLBI can extend this capability to new sources.

\citet{johannsen2012} originally looked at prospects for obtaining
masses of other nearby supermassive black holes using VLBI.  One of
the most promising sources on their list is the Sombrero Galaxy (NGC
4594, M104).  The supermassive black hole in M104 has a mass of $6.6
\times 10^8~M_\sun$ at a distance of $9.9$~Mpc \citep{greene2016},
which leads to a predicted shadow diameter of $\sim 6.8~\mu$as.  At
longer wavelengths, the emission is seen to be very compact, with a
slight elongation indicating the presence of a very weak jet
\citep{hada2013}.  The authors contrast M104 with M87, which has a
much more powerful radio jet, and note the importance of observing
M104 at high angular resolution to determine whether the stark
difference in jet power is due to differences in the black hole spin,
the accretion rate, or other properties of the accretion flow, going
so far as to say that ``M104 and M87 are a unique pair for testing
this issue because the black hole vicinity is actually accessible at a
simlar horizon-scale resolution.''

\begin{figure*}
  \begin{center}
    \resizebox{0.23\hsize}{!}{\includegraphics{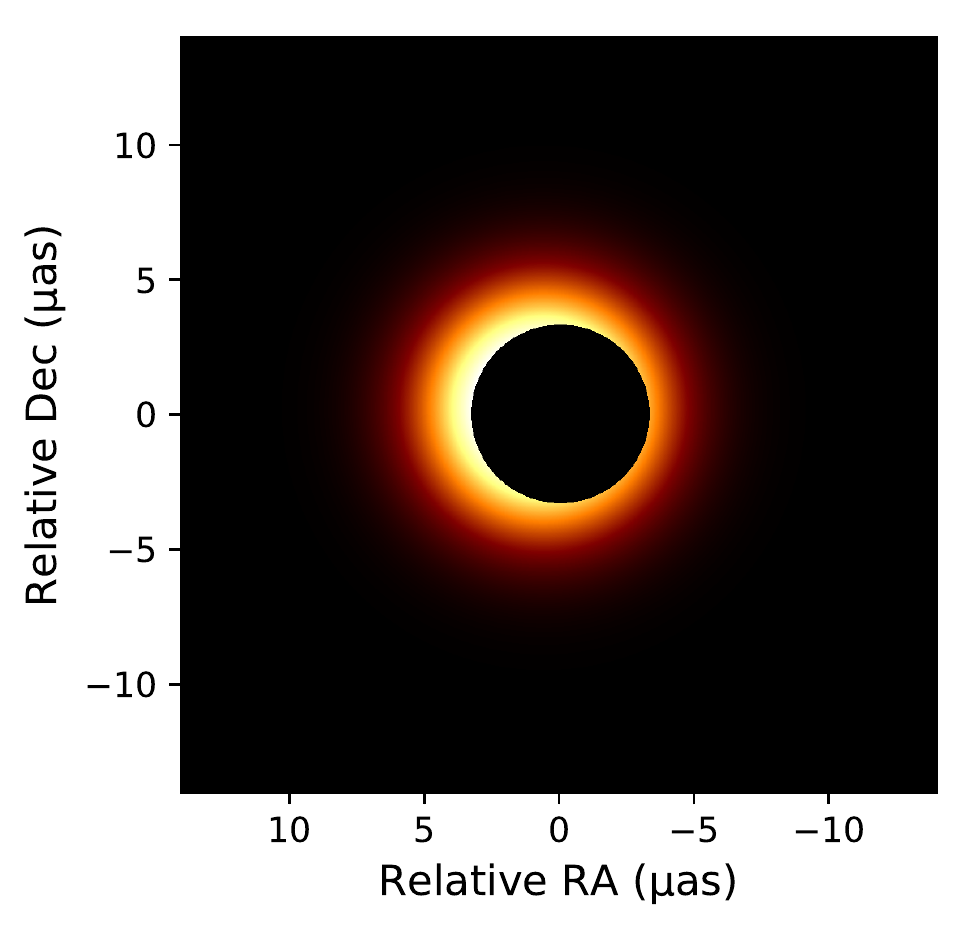}}
    \resizebox{0.23\hsize}{!}{\includegraphics{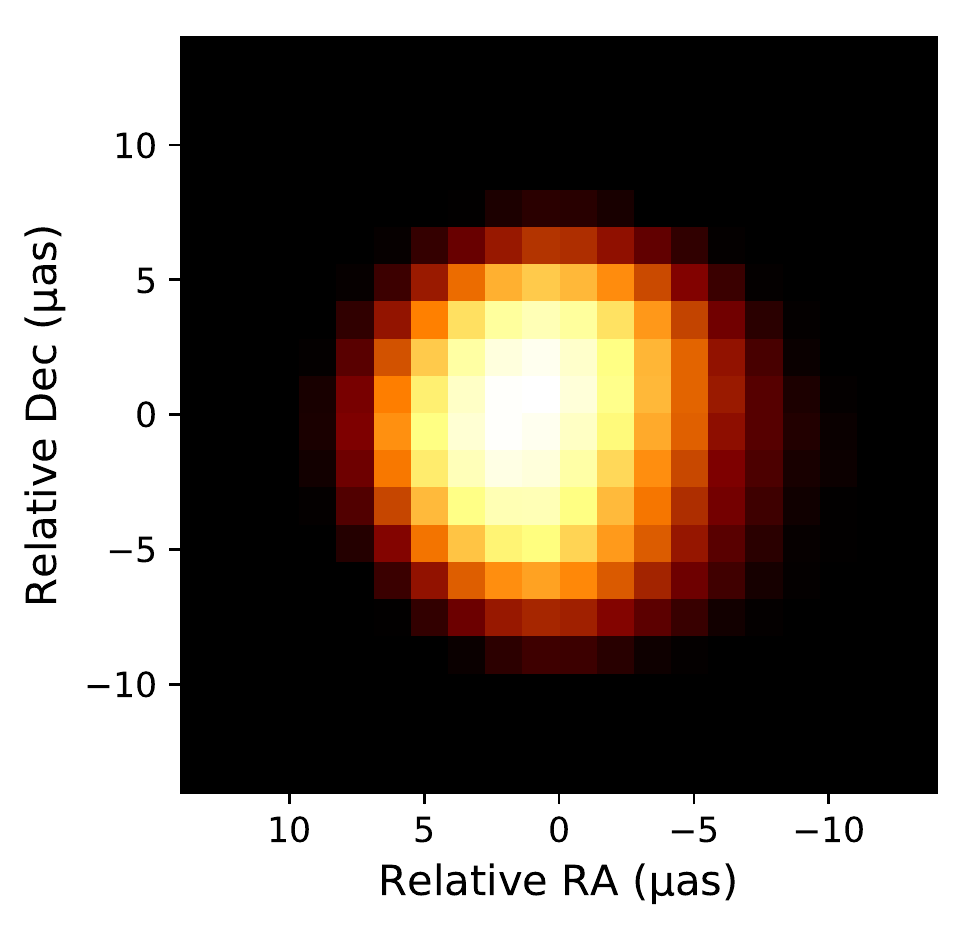}}
    \resizebox{0.23\hsize}{!}{\includegraphics{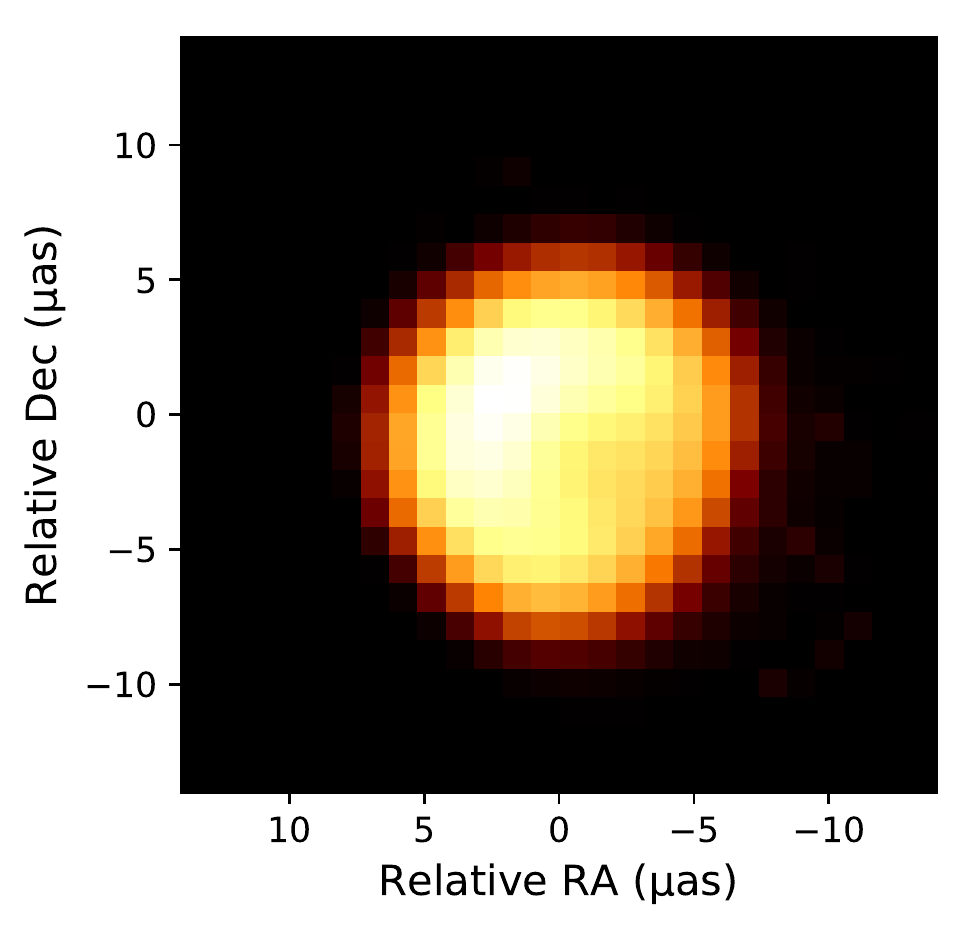}}
    \resizebox{0.23\hsize}{!}{\includegraphics{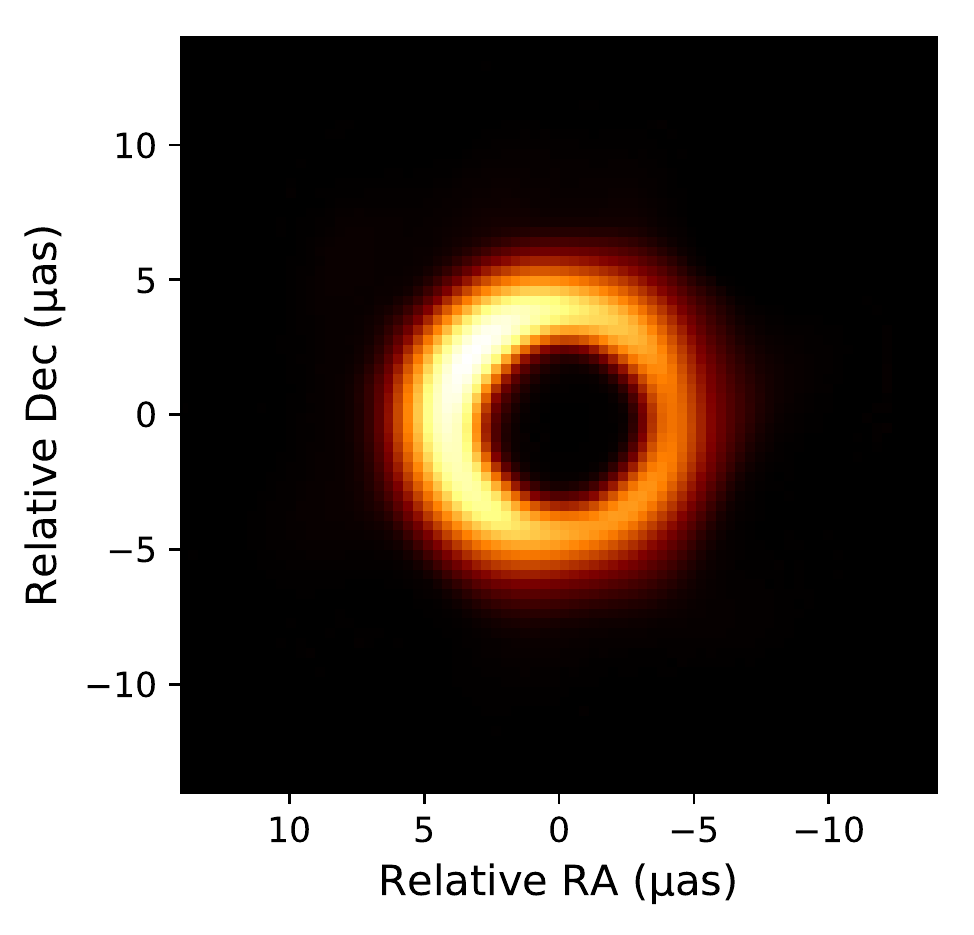}}
    %%%%% For preprint
    %\raisebox{1.75ex}{\resizebox{0.01\hsize}{!}{\includegraphics{f2colorbar.pdf}}}
    %%%%% For print
    \raisebox{2.75ex}{\resizebox{0.01\hsize}{!}{\includegraphics{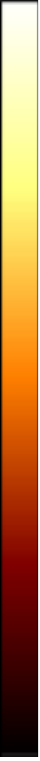}}}
    \caption{Imaging simulation of the Sombrero Galaxy.  \emph{Left to
        right}: Model image, reconstruction with the ground array
      only, with the ground array plus four telescopes in LEO, and
      with the full space array.  The shadow region can be resolved,
      but only if telescopes in MEO/GEO orbits are
      included.  A linear transfer function from zero to the maximum
      pixel value, as shown in the colorbar on the right, is used in
      this and the two subsequent figures.  Pixel sizes are
      automatically selected based on the effective resolution of the
      array.}
    \label{m104-figure}
  \end{center}
\end{figure*}

\begin{figure*}
  \begin{center}
    \resizebox{0.24\hsize}{!}{\includegraphics{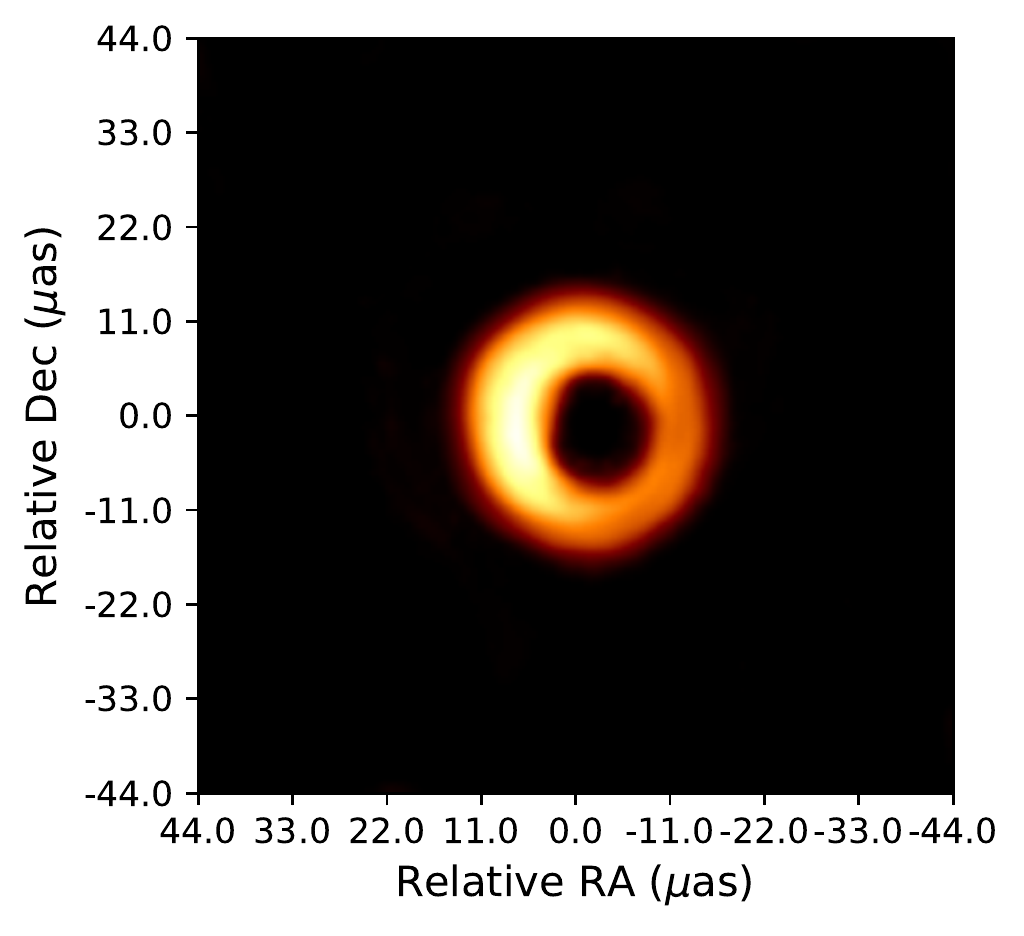}}
    \resizebox{0.24\hsize}{!}{\includegraphics{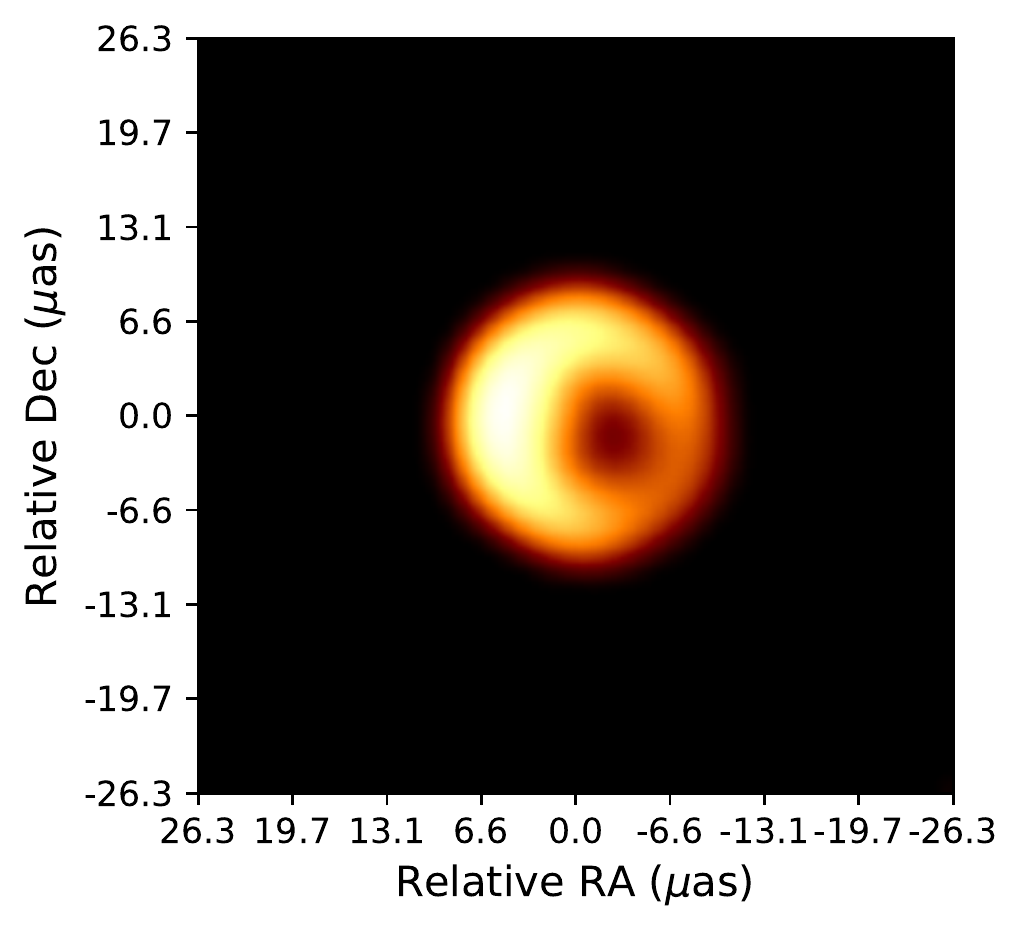}}
    \resizebox{0.24\hsize}{!}{\includegraphics{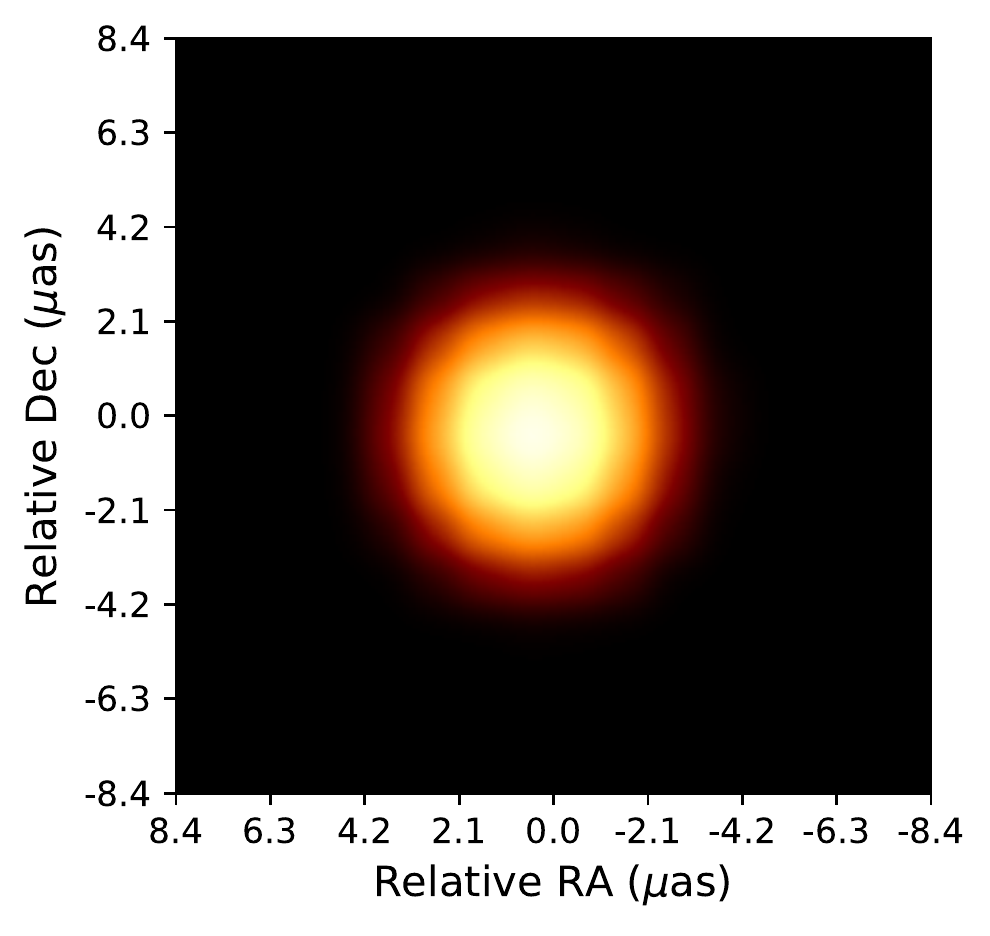}}\\
    \resizebox{0.24\hsize}{!}{\includegraphics{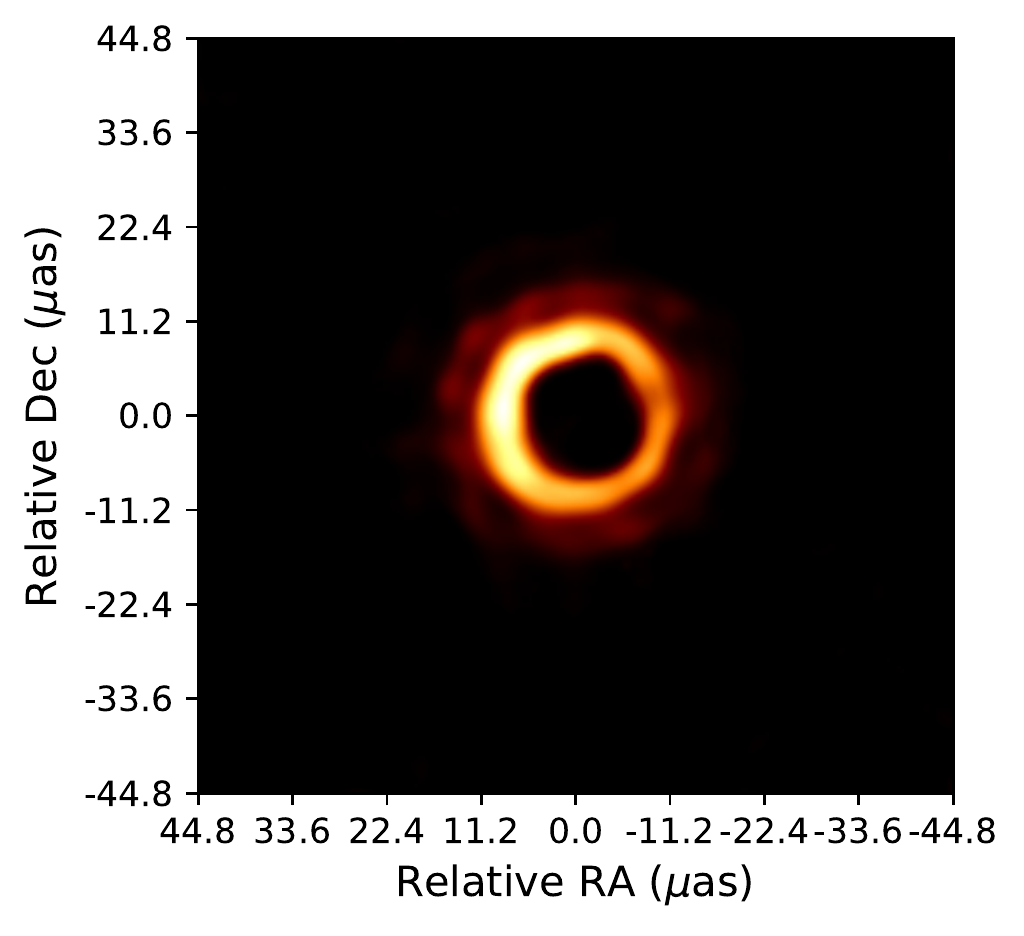}}
    \resizebox{0.24\hsize}{!}{\includegraphics{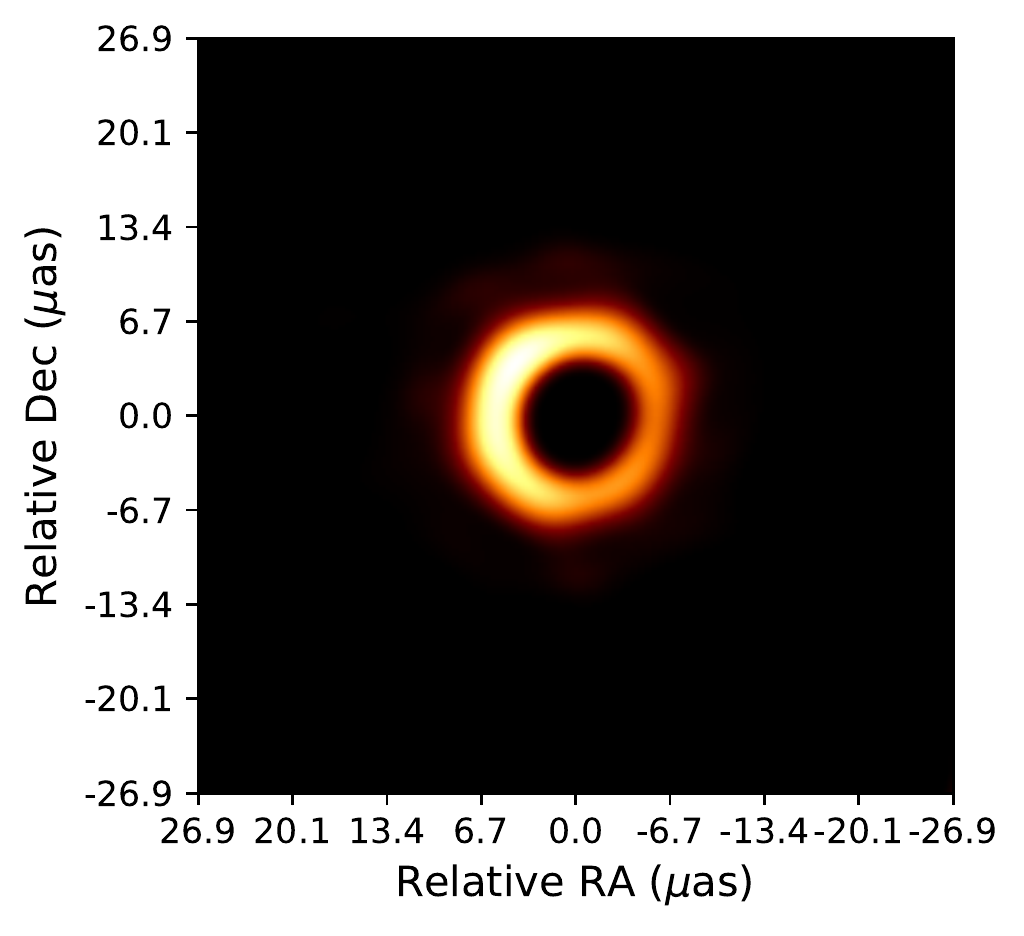}}
    \resizebox{0.24\hsize}{!}{\includegraphics{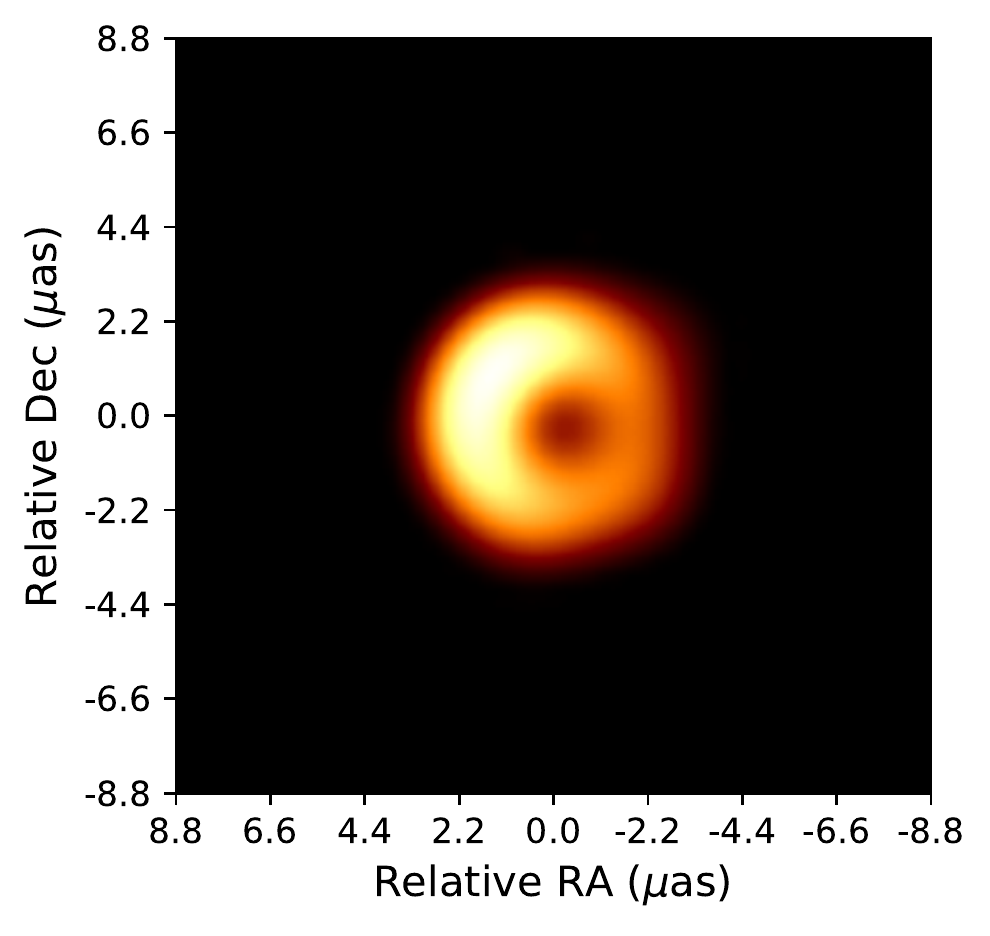}}
    \caption{Test to determine the shadow resolution power of space
      arrays.  In the panels from left to right, the model (left panel
      of Figure~\ref{m104-figure}) was rescaled to have a shadow
      diameter of $15~\mu$as, $9~\mu$as, and $3~\mu$as.  The top row
      shows reconstructions from the ground+LEO array, which can
      marginally resolve shadows down to a diameter of $9~\mu$as.  The
      bottom row shows reconstructions from the full space array,
      which can marginally resolve shadows down to $3~\mu$as.}
    \label{resolution-figure}
  \end{center}
\end{figure*}

Figure~\ref{m104-figure} illustrates the importance of having
telescopes in MEO or GEO orbits to image the black hole region of the
Sombrero Galaxy.  Neither the ground array alone nor an array
consisting of both ground and LEO telescopes is sufficient to resolve
the black hole shadow region.  The shadow is well imaged when MEO/GEO
satellites are included.  (We have assumed a static image for these
simulations, but several tracks may be required if the source is in an
active state, since $t_g$ is about an hour for the Sombrero Galaxy.)
Simulations demonstrate that the full space array could resolve a
black hole shadow down to approximately $3~\mu$as in diameter
(Fig.~\ref{resolution-figure}), which would add M104, IC~1459, M84,
and perhaps IC~4296 to the list of supermassive black holes that are
bright enough and are predicted to have a sufficiently large shadow to
be resolved\footnote{The very bright source Cen~A may also be on this
  list, although the black hole mass estimate from stellar kinematics
  is less encouraging \citep{cappellari2009}.} (see
\citealt{johannsen2012} for mass and distance estimates).

\subsection{M87 Jet Launch}

The shadow of M87 is within the reach of the resolution of a
ground-based array alone, and it is probable that the EHT will produce
successful images of the M87 shadow within the next few years.  Models
of M87 suggest that the 1.3~mm emission is mainly concentrated near
the shadow region
\citep[e.g.,][]{broderick2009,dexter2012,moscibrodzka2016}, a
conclusion supported by early EHT data \citep{doeleman2012}.  In
contrast, longer-wavelength data show a prominent jet extending far
away from the location of the black hole \citep[e.g.,][]{walker2018}.
How this jet is launched is an open question.

\begin{figure*}
  \begin{center}
    \resizebox{0.24\hsize}{!}{\includegraphics{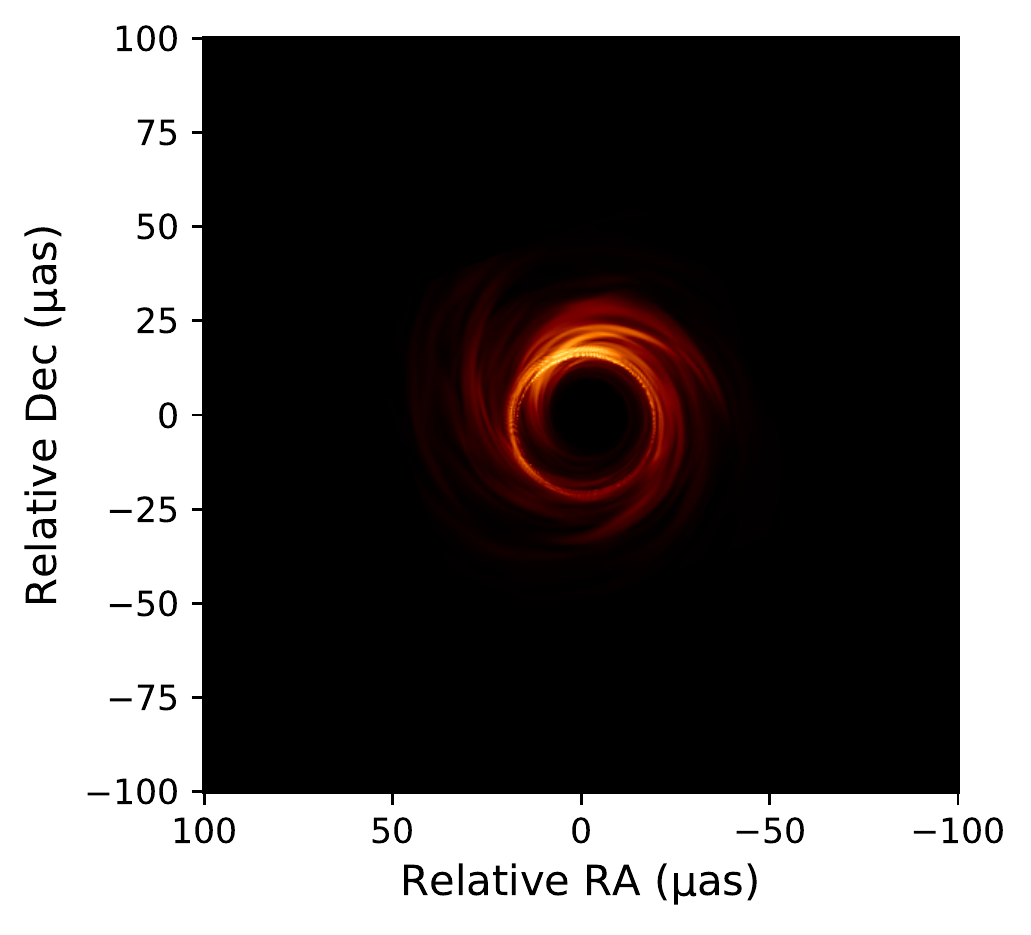}}
    \resizebox{0.24\hsize}{!}{\includegraphics{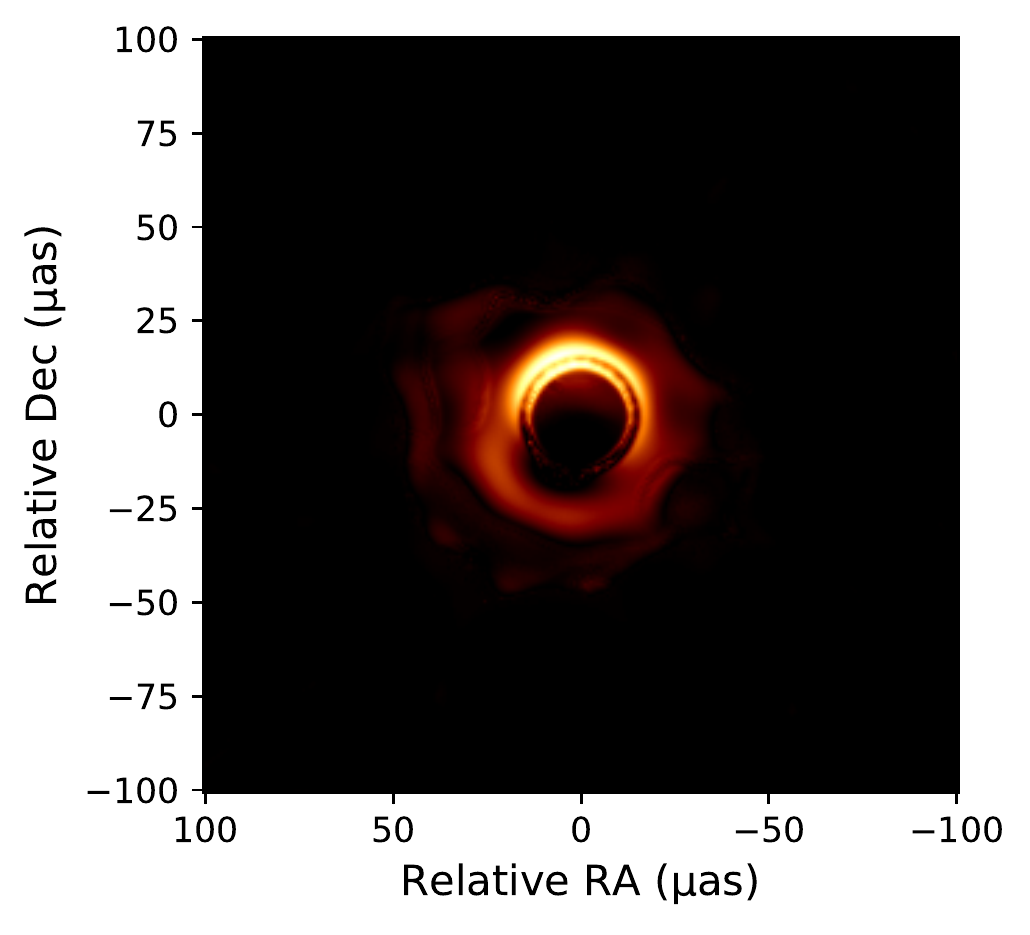}}
    \resizebox{0.24\hsize}{!}{\includegraphics{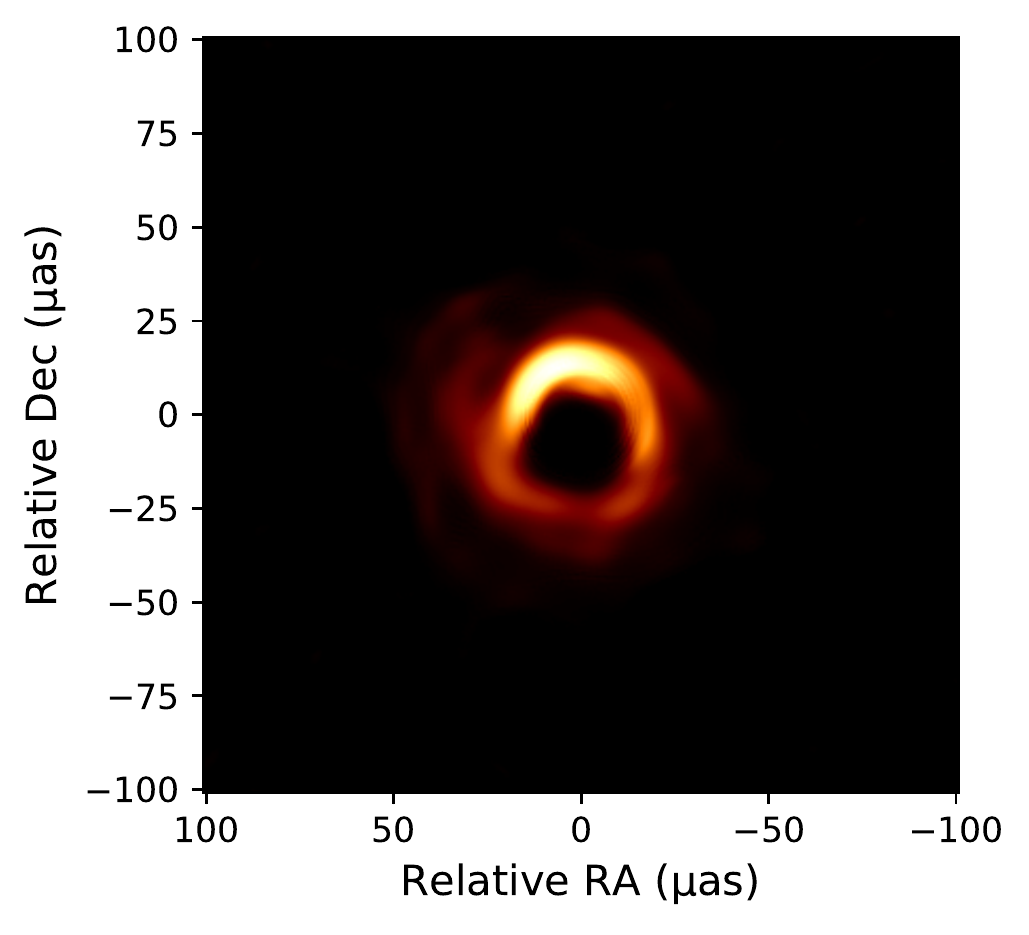}}
    \resizebox{0.24\hsize}{!}{\includegraphics{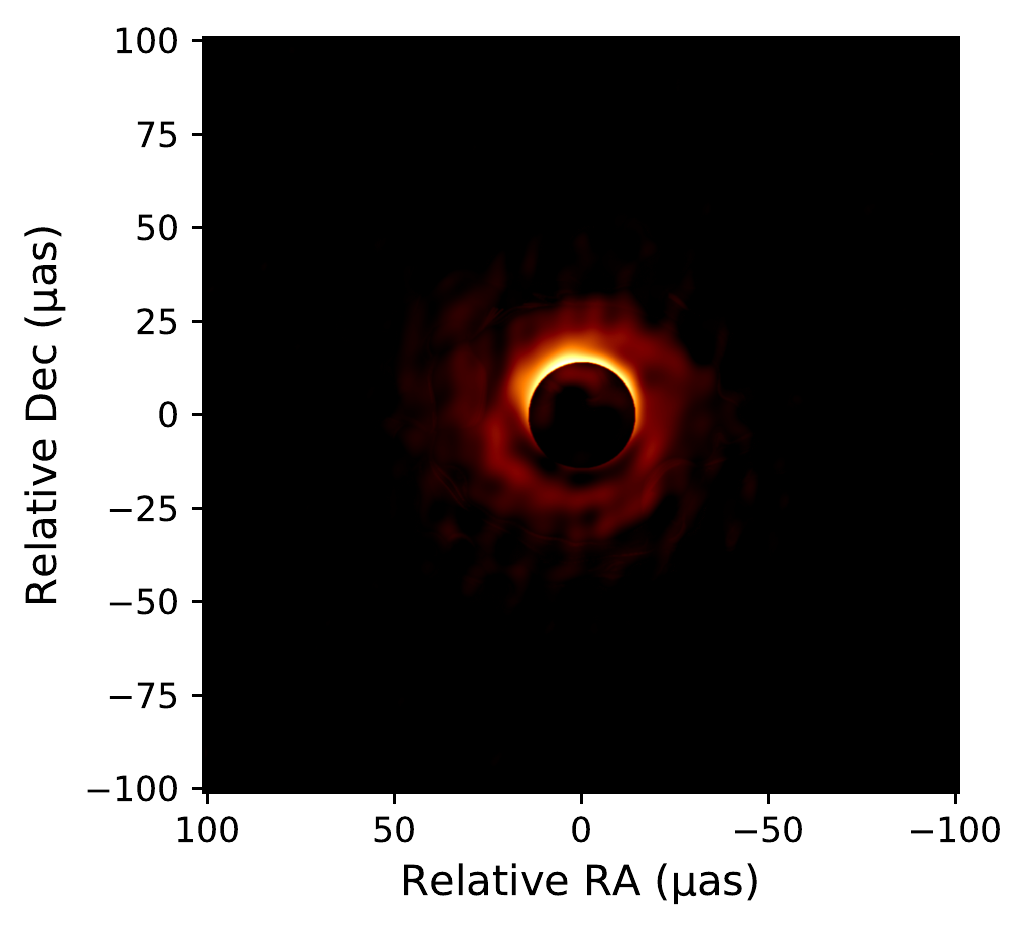}}
  \end{center}
  \caption{\emph{Left to right}: Model of the M87 accretion disk and
    jet \citep{moscibrodzka2016,moscibrodzka2017} along with
    reconstructions of simulated data from the ground-only,
    ground+LEO, and full space-VLBI array.  The helical structure,
    which is barely resolved with LEO satellites, is clearly visible
    when MEO/GEO satellites are included.}
  \label{m87-figure}
\end{figure*}

Inhomogeneities in the accretion disk and jet may produce useful
tracers of the motions and magnetic field in the jet launch region
around the black hole.  The models of
\citet{moscibrodzka2016,moscibrodzka2017} illustrate physically
plausible emission profiles that might be seen at millimeter
wavelengths.  Imaging and tracking the evolution of these
substructures may help to distinguish whether the observed emission is
associated with the jet sheath and whether the disk and jet have
different proton-to-electron temperature ratios.  A space-VLBI array
would provide greater clarity than is available from ground-based VLBI
alone (Fig.~\ref{m87-figure}).

\subsection{AGN Jet Collimation and Variability}

Higher-resolution, higher-fidelity imaging would also be a boon to
studies of AGN on spatial scales much greater than $r_g$.  In the last
decade, multiwavelength observations of AGN have been a subject of
active investigation.  In particular, after the launch of the Fermi
telescope and the advent of ground Cherenkov telescopes (e.g., HESS,
MAGIC, and VERITAS), VLBI has played an important role in locating the
flaring counterpart of high-energy emission \citep{marscher2008}.
High angular resolution observations at millimeter VLBI wavelengths
have been useful for constraining properties of the flaring region
such as the source size \citep{akiyama2015}.  In the context of
multi-messenger observations, resolving the detailed structure of jets
will be increasingly important in the next decades, including
next-generation Cherenkov telescopes and neutrino observatories.

\begin{figure*}
  \begin{center}
    \resizebox{0.24\hsize}{!}{\includegraphics{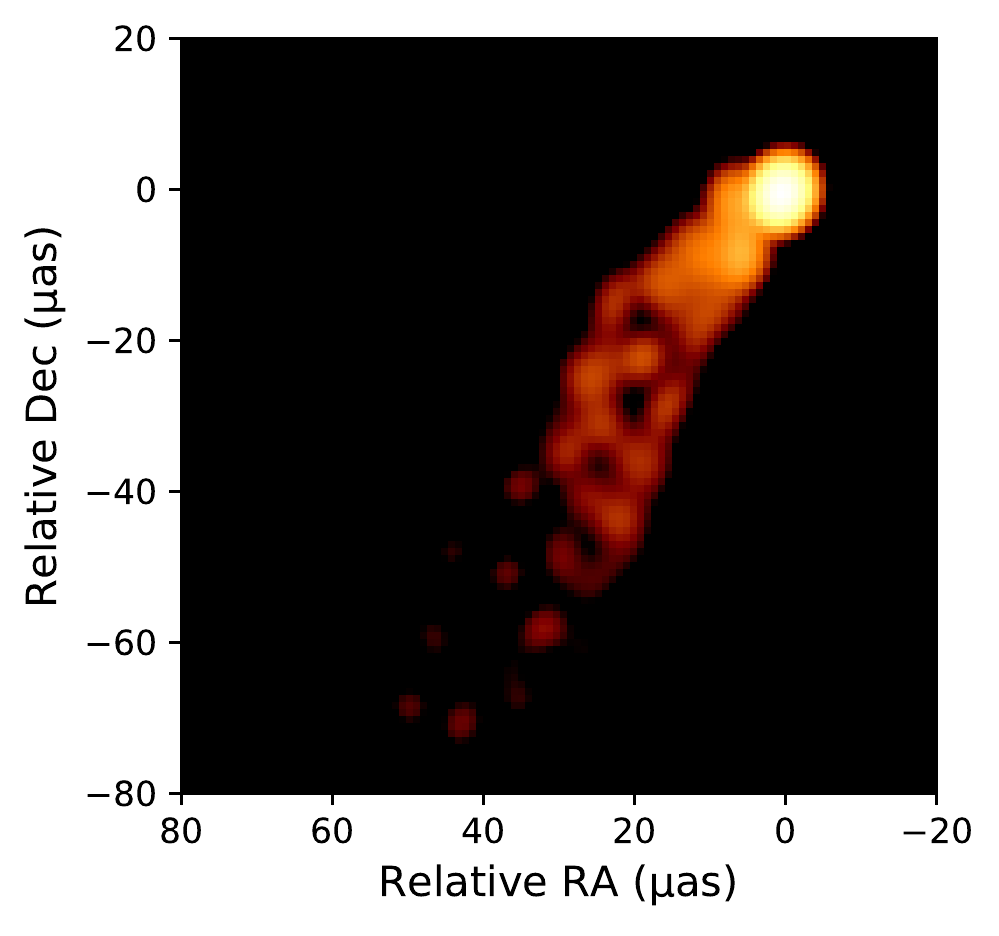}}
    \resizebox{0.24\hsize}{!}{\includegraphics{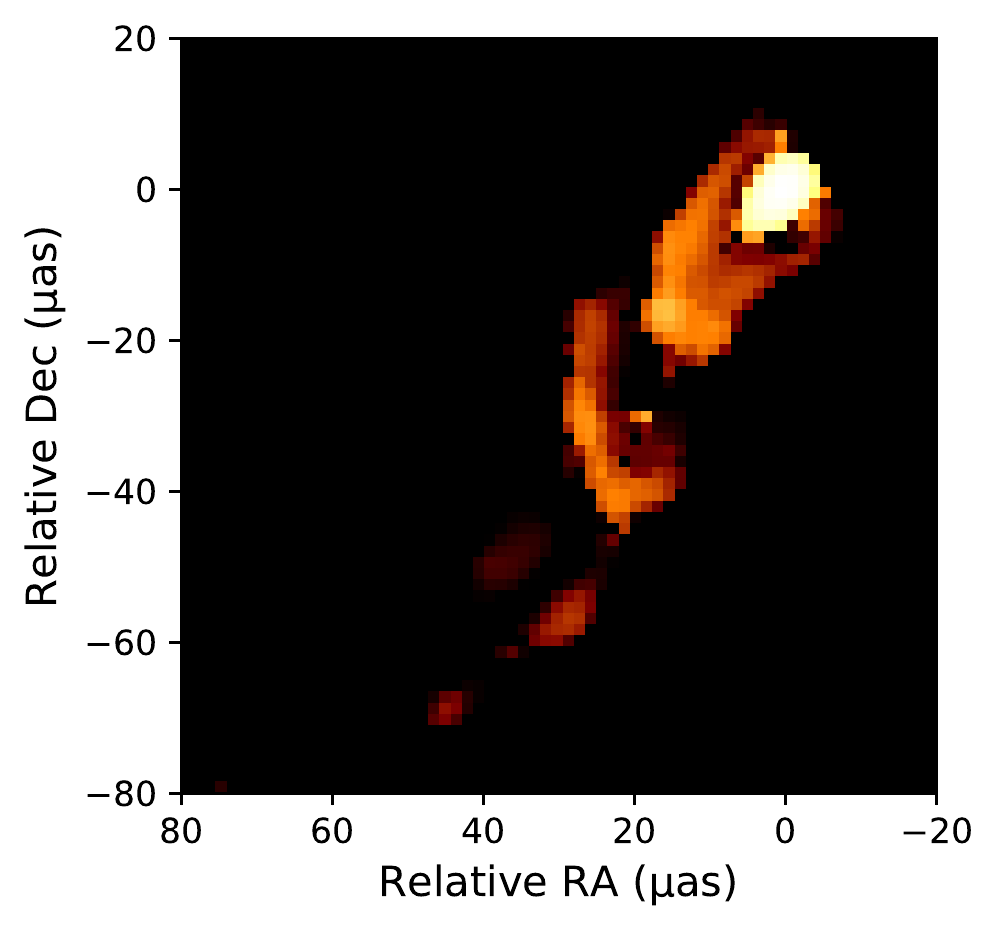}}
    \resizebox{0.24\hsize}{!}{\includegraphics{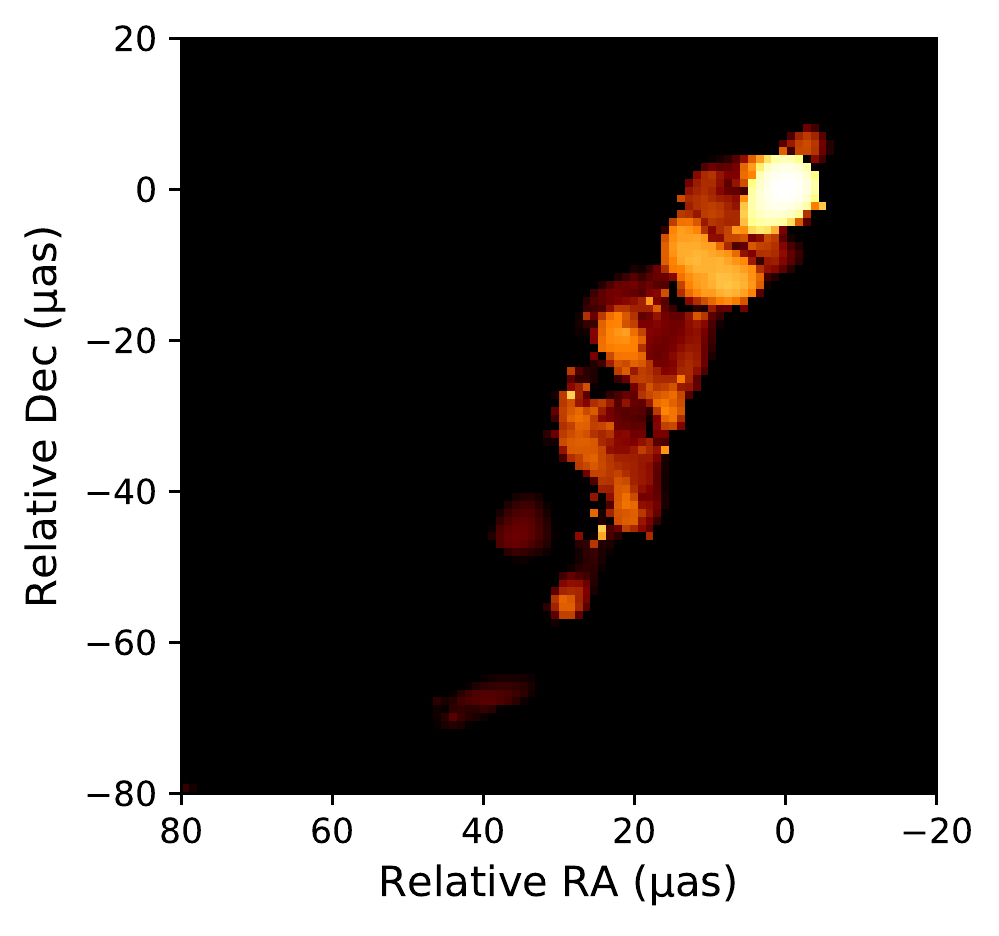}}
    \resizebox{0.24\hsize}{!}{\includegraphics{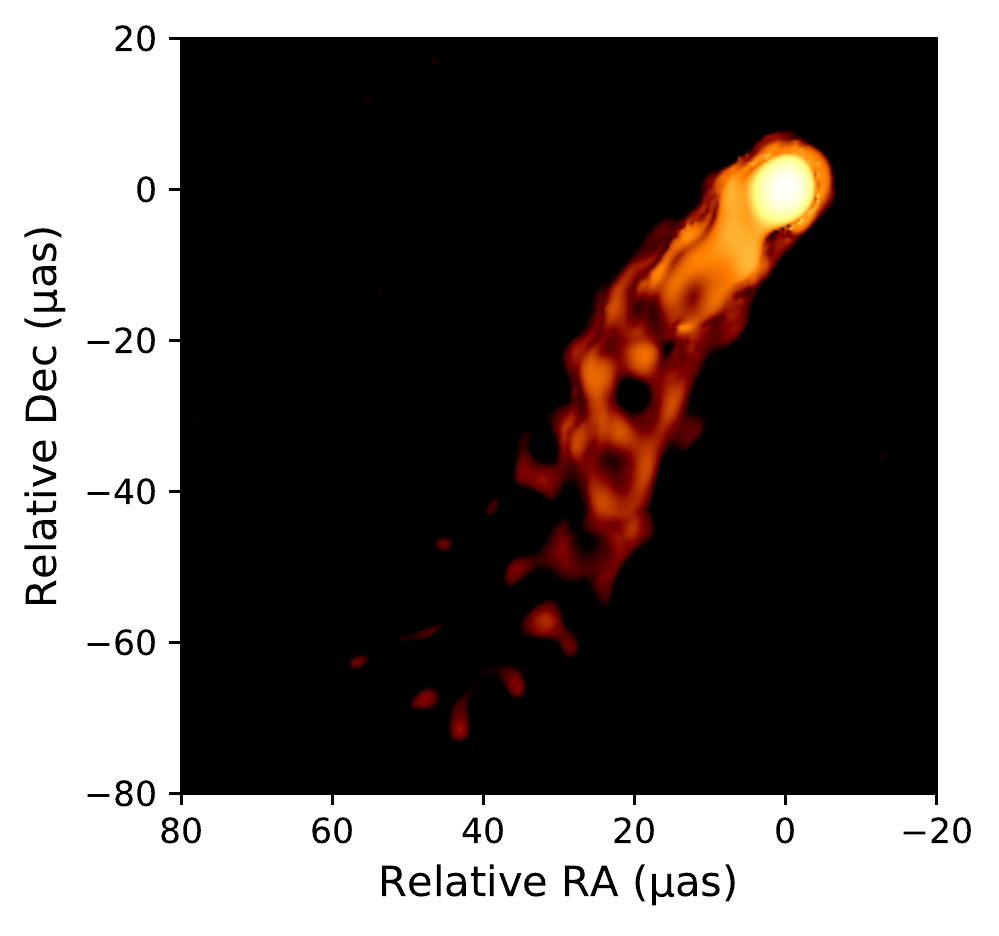}}
    \caption{\emph{Left to right}: Model of the Mrk~501 jet based on
      rescaling the MOJAVE (Monitoring Of Jets in Active galactic
      nuclei with VLBA Experiments) 15~GHz image \citep{lister2018}
      along with reconstructions of simulated data from the
      ground-only, ground+LEO, and full space-VLBI array.  The
      increased resolution of the full space array is required to
      provides a much truer reconstruction of the details of the jet,
      including the narrow opening angle by the core, the
      limb-brightening of the jet, and faint substructures within the
      jet.  A logarithmic transfer function, with the color range
      spanning three orders of magnitude in dynamic range, is used to
      highlight weak features in the jet.}
    \label{mrk501-figure}
  \end{center}
\end{figure*}

Increased resolution would also allow AGN jet profiles to be traced
closer to the black hole.  While radio galaxies often exhibit
parabolic collimation profiles near the black hole and a transition to
a more conical profile outside
\citep{asada2012,nagai2014,boccardi2016,tseng2016,giovannini2018,hada2018,nakahara2018},
quasar jet properties are less well studied.  A study of the 3C~273
jet is suggestive of a similar transition near its Bondi radius,
possibly indicating that jet collimation processes are universal
\citep{akiyama2018}.  It has been difficult to obtain enough data to
test this hypothesis due to a lack of sufficient resolution.
Figure~\ref{mrk501-figure} illustrates that a full space-VLBI array
may be necessary to accurately determine jet collimation, with
lower-resolution arrays possibly providing an incorrect qualitative
understanding of the collimation near the core.

\section{Discussion}

In this work, we have explored a space-VLBI concept that includes
space--ground baselines.  Such an array can provide fast and/or dense
$(u,v)$ coverage with only a few orbiters.  Large apertures on the
ground provide a cost-effective way to maximize sensitivity.  Our
simulations demonstrate that a modest space-VLBI array is sufficient
to make significant breakthroughs, including measuring the shadows of
a larger sample of supermassive black holes, providing detailed images
of the jet launch region in M87, and resolving the collimation
profiles of a larger collection of AGN jet sources near the black
hole.

While many architectural decisions for a space-VLBI concept can be
postponed, the question of whether to include space--ground baselines
or rely upon only space--space baselines is fundamental and must be
decided early.  Bringing the data back to the ground has some key
advantages, including much faster $(u,v)$ coverage, better
sensitivity, and greater robustness in fringe finding, since a
correlator on the ground can more easily handle uncertainties in
spacecraft orbits and local-oscillator frequencies/timing.  An
underappreciated advantage of building space--ground VLBI into the
architecture is extensibility.  For instance, a pair of telescopes
that are in identical orbits but for a small vertical offset, as in
the Event Horizon Imager (EHI) concept \citep{roelofs2018} could
easily be accommodated in a space--ground architecture by the simple
addition of a second satellite in MEO.  The reverse statement is not
true; it would be extremely difficult, if not impossible, to
incorporate a number of additional satellites (and ground stations)
into the EHI architecture.

For the purposes of this study, we have limited consideration to only
the 230~GHz (1.3~mm) band.  Nevertheless, as is clear from many of the
other articles in this issue, there is a strong scientific case for
including other frequency bands, especially at longer wavelength.
Indeed, some aspects of the science in this work (e.g., jet
collimation studies) would benefit also benefit from
higher-resolution, multiwavelength observations at centimeter and
millimeter wavelengths.  Having multiple observing bands on VLBI
satellites is a cost-effective way to increase the science per dollar
as well as the observing duty cycle, since weather conditions will not
always be suitable for 230~GHz observing from the ground.

\acknowledgments

This material is based upon work supported by the National Science
Foundation (NSF) under grant numbers AST-1440254, AST-1614868, and
AST-1659420.  M.~S.\ acknowledges support from the NSF Research
Experiences for Undergraduates program.  K.~A. acknowledges support
from the Jansky Fellowship of the National Radio Astronomy Observatory
(NRAO), a facility of the NSF operated by Associated Universities,
Inc.  This research has made use of data from the MOJAVE database that
is maintained by the MOJAVE team \citep{lister2018}.  We thank Michael
Hecht for valuable discussions regarding the technical landscape of
satellite systems and Daniel Palumbo for discussions relating to
space-VLBI concepts with a small number of antennas in LEO.

\software{ehtim (\url{https://achael.github.io/eht-imaging}), SMILI
  (\url{https://github.com/astrosmili/smili})}

\end{document}